\documentclass[letterpaper, 10 pt, conference]{ieeeconf}  % Comment this line out if you need a4paper

\IEEEoverridecommandlockouts                              % This command is only needed if 
\usepackage[noadjust]{cite}
\usepackage{amsmath,amssymb,amsfonts}
\usepackage{nccmath}
\usepackage{algorithmic}
\usepackage{graphicx}
\usepackage{textcomp}
\usepackage[usenames,dvipsnames,svgnames,table]{xcolor}
\usepackage{cuted}
\usepackage{setspace}
\usepackage{hyperref}
\usepackage{xr}

\usepackage[normalem]{ulem}
\newcommand\redsout{\bgroup\markoverwith{\textcolor{red}{\rule[0.5ex]{2pt}{0.4pt}}}\ULon}

\usepackage{amsthm}
\theoremstyle{definition}
 
\newtheorem{theorem}{Theorem} 
\newtheorem{lemma}[theorem]{Lemma}

\hypersetup{
	colorlinks=true,       % false: boxed links; true: colored links
	linkcolor=blue,        % color of internal links
	citecolor=blue,        % color of links to bibliography
	filecolor=magenta,     % color of file links
	urlcolor=blue         
}

\newcommand{\Max}[1]{\raisebox{0.6ex}{\scalebox{0.8}{$\displaystyle \max_{#1}\;$}}}

\title{\LARGE \bf
	Towards Stable Interstellar Flight:\\
	Levitation  of  a  Laser-Propelled Sailcraft*}

\author{Afroza Shirin$^{1}$, 
	 Edl Schamiloglu$^{2}$, %Cornel Sultan$^{2}$, Yin Yang$^{3}$, 
	and Rafael Fierro$^{2}$%<-this % stops a space
	\thanks {*This work is supported by Breakthrough Starshot Foundation LLC.}
	\thanks{$^{1}$Aerospace and Mechanical Engineering, University of Texas at El Paso, El Paso, Texas, USA. {\tt\small E-mail: ashirin@utep.edu}}%
	\thanks{$^{2}$Electrical \& Computer Engineering, University of New Mexico, Albuquerque, NM, USA. {\tt\small E-mail: \{edls, rfierro\}@unm.edu}}%{\tt\small E-mail: \{ashirin, edls, rfierro\}@unm.edu}}
}

\begin{document}

	\maketitle
	\thispagestyle{empty}
	\pagestyle{empty}
	%\pagenumbering{gobble} % will not show the page number

	%%%%%%%%%%%%%%%%%%%%%%%%%%%%%%%%%%%%%%%%%%%%%%%%%%%%%%%%%%%%%%%%%%%%%%%%%%%%%%%%
	\begin{abstract} 
	Exploring and traveling to distant stars has long fascinated humanity but has been limited due to the vast distances. The Breakthrough Starshot Program aims at eliminating this limitation by traveling to Alpha Centauri, which is 4.37 light-years away. This is only possible if a vehicle travels at a substantial fraction of the speed of light. The Breakthrough Starshot Program initiative is to develop a proof-of-concept that is accelerating a sail to relativistic speeds using a laser beam aimed at the sail. At this high speed, while stable beam riding is one of the crucial concerns of this concept, the dynamic stability analysis of a sail is hardly present in the previous literature. Furthermore, it is important to investigate the dynamic stability in the experiment before driving the sail to relativistic speeds. As a proof-of-concept, we study the dynamic stability of the sail levitated at a certain height by a laser beam.	The sail's dynamics are modeled as a rigid body whose shape is parameterized by a sweep function. We estimate the region of attraction (ROA) for dynamic stability analysis using Lyapunov theory and sum-of-square (SOS) programming. The ROA confirms how many transverse and angular perturbations a levitated sail can tolerate. We also conclude on some of the important parameters of the sail that affects the dynamic stability. Simulation results validate our theoretical analysis.

	% Currently, there is no dynamic stability analysis present and no agreement on the proper shape of the LightSail in the literature. Therefore, to study the dynamic stability analysis of the LightSail, we model the sail as a 3D rigid body. We also include a generalized sail shape as a parameterized sweep function. In this paper, we add self-stabilizing dynamics. We anticipate that the technological advances in materials and optomechanics would allow us to induce a dynamic damping effect. In this paper, we model the dynamics so that the sail will have a self-stabilizing effect. We then analyze the stability of the LightSail by using Lyapunov theory. Simulation results validate our theoretical analysis.
	
\end{abstract} 

	
	\section{Introduction}
	The Breakthrough Starshot project aims to demonstrate a proof-of-concept  for ultra-fast laser beam-driven sail and lays the foundations for the first launch to Alpha Centauri within the next generation \cite{finkbeiner2017near,daukantas2017breakthrough,parkin2018breakthrough}. The idea is to propel a sail, which will carry ultra-low-mass ``starchips" to 20\% of the speed of light. The  spacecraft would spend 20 years getting to Alpha Centauri and another 4.25 years sending back	its data so that they can be analyzed within next generation. 
	%In the last decade, rapid technological advances have opened up the possibility of laser-driven sail travel at a significant fraction of the speed of light so that the sail can arrive at Alpha Centauri within the lifetime of those who launched it \cite{savu2020structural,bandutunga2021photonic,deri2011semiconductor,worden2021progress}.	
	The stability of the sail is a key consideration in the success of this concept and is affected by sail shape, laser beam profile, beam jitter, mass distribution, payload, spinning dynamics, flexibility, material properties, etc.   
    There are several investigations on the stability of a sail based on the sail configuration \cite{chahine2003dynamics,benford2002experimental,popova2017stability,manchester2017stability,genta2017preliminary}. 
    There are some other works where the material and optical properties have been included to investigate the stability \cite{siegel2019self, ilic2019self,chu2021parametric,salary2020photonic,gao2021restoring}. 
	
	Although these initiatives towards stability analysis of the sail are promising,  they are based on static stability analysis and carry some limitations.  First, static stability, which is the linearization around an equilibrium, can be used to indicate stability/instability for the particular parameters they used, but the conclusions may not apply to the parameters or timescales other than those considered in the study. Second, the precise nano-patterning and handling of a lightweight macroscopic substrate is a delicate task. Also, the loading of the object in the laser beam will be difficult as it will require the system to be released in a very controlled way. 
	
	As the sail itself does not have any propeller or control and the laser will be ground based or in a low earth orbit, the perturbations in translational and angular positions with respect to the laser beam are very critical and measuring the maneuverability is very important. This analysis can be done by using dynamic stability or nonlinear Lyapunov stability \cite{khalil2002nonlinear}, which has barely made an appearance in the  current sail literature.  One of the aspects of dynamic stability analysis is to estimate the region of attraction (ROA). Any perturbation of the states within the ROA will converge to the equilibrium point.  The ROA can be used to quantify the maneuverability with respect to the laser beam. However, estimation of the ROA of an equilibrium point for uncertain systems is itself a challenging problem that requires  the development of  computationally efficient methods. 
	To address the dynamic stability analysis of a levitated sail, we estimate the ROA.
	In this article, we consider the sail dynamics as a 3D rigid body. We parameterize the sail shape. Then we use Lyapunov theory and sum-of-squares (SOS) programming to estimate the ROA \cite{parrilo2000structured}. Finally, we analyze how the ROA is affected as a sail's parameter changes.

	\section{Parameterized Sail}
\begin{figure}[h]
     	\centering
     	\includegraphics[scale = 0.69]{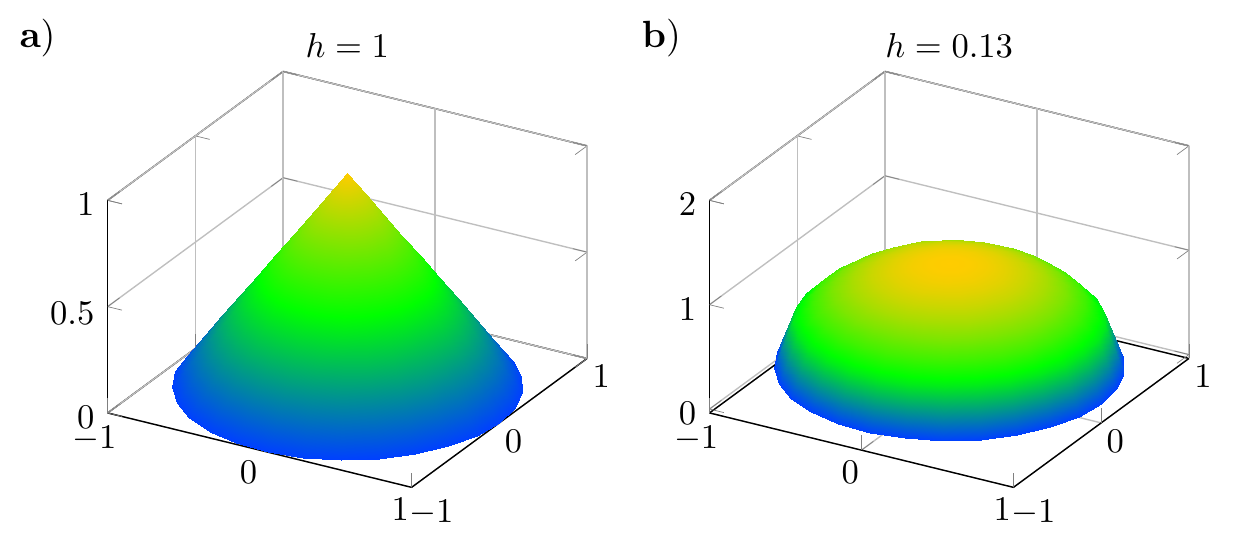}
     	\vspace{-0.5 cm}
     	\caption{a) A conical sail with base radius $R = 1$, $h = 1$ and $\alpha = \frac{\pi}{4}$. b) A portion of a spherical-cap sail with base radius $a =0.5$, $h=0.13$ and with a radius of curvature $R=1$.}
         \label{fig:quadratic}
    \end{figure}

	The whole vehicle consists of a sail, a mast, and a payload. The payload is assumed to be attached by the mast to the sail. To generate the sail surface a parameterized sweep curve in the x-z plane is rotated about the z-axis \cite{shirin2021modeling}. In this article, two types of sail shapes are considered: 1) conical sail and 2)  spherical-cap sail. The conical sail is studied in \cite{chahine2003dynamics,manchester2017stability} and a spherical-cap sail in  \cite{popova2017stability}. For the conical sail, the following curve is used
	\begin{equation}\label{eq:polynomial_sail}
		%		\begin{medsize}
		%			z = g(x) = c_0 + c_1 (\frac{x}{R}) + c_2 (\frac{x}{R})^2 +c_3 (\frac{x}{R})^3 + c_4 (\frac{x}{R})^4.
		g_s(x) = c_0 + c_1 \left(\frac{x}{R}\right) + c_2 \left(\frac{x}{R}\right)^2 +c_3 \left(\frac{x}{R}\right)^3 + c_4 \left(\frac{x}{R}\right)^4.
		%		\end{medsize}
	\end{equation}
	Here $c_0, c_1, c_2, c_3$, and $c_4$ are coefficients of the polynomials and define different shapes of the sail with base radius $R$. 	A conical sail with base radius $R$ and cone angle $\alpha$ can be obtained by setting $c_0 = R\tan(\alpha), c_1 = -R\tan(\alpha), c_2 = c_3 = c_4 = 0$ in Eq.\ \eqref{eq:polynomial_sail}. Here the angle $\alpha$ is between the conical sail surface and the horizontal plane. For the conical sail, the cone angle $\alpha$ is considered an important parameter and it is analyzed how the dynamical stability is affected as the parameter $\alpha$ changes. 
	To generate a spherical-cap sail, the following function is used
	\begin{equation}\label{eq:spherical_sail}
		g_s(x) = \sqrt{R^2 - x^2}-\sqrt{R^2 -a^2}.
	\end{equation}
	Notice that the mast length is an important parameter to analyze the dynamic stability. A conical sail and a spherical-cap sail are presented in Fig \ref{fig:quadratic}.
\section{Modeling the Dynamics of the Sail}\label{sec:model}
	For mathematical modeling, we assume that the whole vehicle, which includes a sail, a mast, and a payload, is a rigid body. The sail shape is chosen from Section II, the length of the mast is $l$, and the total mass of the vehicle is $m$. The sail is fully reflective and not a multi-reflector \footnote[1]{multi-reflector: A surface that reflects the light more than once.}. 
	The vehicle motion is represented as a translation of the vehicle's center of mass (CM) and a rotation of the body frame ($x_b-y_b-z_b$) with respect to the inertial frame ($x-y-z$). The vehicle's CM position is described by its inertial Cartesian coordinates, $\textbf{r} = (x, y, z)$, and its orientation with respect to the inertial reference frame by three Euler angles $\boldsymbol{\alpha}_E = (\psi , \theta, \phi)$. The convention of rotations is considered as  $ x-y'-z''$ by following the sequence: (1) rotation about the x-axis by angle $\phi$; (2) rotation
	about the new position of the y-axis by angle $\theta$; (3) rotation about the new position of the	z-axis by angle $\psi$. Now the vehicle dynamics is described by 12 ODEs \cite{hughes2012spacecraft}: 
	\begin{subequations}\label{eq:dynamics}
		\begin{align}
		\dot{\textbf{r}} & = 
		\textbf{v}, \\
		\dot{\boldsymbol{\alpha}}_E & = L^I_B \boldsymbol{\omega}_b\label{eq:LIB}, \\
		\dot{\textbf{v}} & = \frac{\textbf{F}}{M} + \textbf{g}, \label{eq:dynamics_c}\\
		\dot{\boldsymbol{\omega}}_b & = J^{-1} [- \boldsymbol{\omega}_b^{X} \times J\boldsymbol{\omega}_b + \boldsymbol{\tau}_b], \label{eq:dynamics_d}
		\end{align}
	\end{subequations}
	where 
	%vector $ \textbf{r}= \{x, y, z\}$  denote the inertial coordinates of the  CM of the vehicle, 
	$\textbf{v}$ is the inertial velocity of the vehicle,
	$\boldsymbol{\omega}_b^{X}$ is the skew-symmetric matrix associated with the vehicle's angular velocity vector,
	%$\boldsymbol{\alpha}$ is the angular displacement in the inertial frame, 
	$\boldsymbol{\omega}_b$ is the angular velocity of the vehicle in the inertial frame and expressed in the body-frame, $J$ is the inertia matrix with respect to the vehicle's CM, $\textbf{g} = [0,0,-9.8]$ m/s is the gravity vector, $M$ is the total mass of the vehicle. In this formulation, we assume that the origin of the body axis moves along the inertial $z$-axis. The force $\textbf{F}$ and the torque $\boldsymbol{\tau}_b$ are the net external force and torque acting on the body, respectively. When the laser beam is radiated onto the sail, the force and the torque are given by
	\begin{equation}\label{eq:force}
	\begin{aligned}
		\textbf{F}(\textbf{x},\Theta) = \left(R_I^B\right)^{-1} & \left[ \iint_{S(\Theta_1)} 2 \frac{ P(\textbf{x},\Theta_2) \hat{\textbf{b}}\cdot \hat{\textbf{n}}(\textbf{x}) }{c} \hat{\textbf{n}_b} dS \right] ,  \\  
		& \quad \hat{\textbf{n}}(\textbf{x}) = \left(R_I^B\right)^{-1}  \hat{\textbf{n}_b}. 
	\end{aligned}
	\end{equation}

	\begin{equation}\label{eq:torque}
	\boldsymbol{\tau}_b(\textbf{x},\Theta)  = \iint_{S(\Theta_1)} 2 \frac{ P(\textbf{x},\Theta_2) \hat{\textbf{b}} \cdot \hat{\textbf{n}}(\textbf{x}) }{c} (\textbf{r}_b(\textbf{x}) \times \hat{\textbf{n}}_b) dS,
	\end{equation}
	where $c$ is the speed of light, $S$ is the domain of integration of the surface of the sail, $\hat{\textbf{n}}_b$
	is the unit vector normal to the sail surface at the generic point on the sail, $\hat{\textbf{b}}$ is a unit vector parallel to the beam axis, $\textbf{r}_b$ is the position vector from the vehicle CM to the point $\textbf{x}$ in the body frame. Here $\Theta = \{\Theta_1,\Theta_2\}$, $\Theta_1$ is the set of parameters used to describe the sail geometry, and $\Theta_2$ is the set of parameters to express the beam profile. The laser beam is assumed to have a radially symmetric Gaussian power distribution with full-width-at-half-maximum (FWHM)  $w = 2 \sigma \sqrt{2 \ln(2)}$, where  $\sigma$ is the standard deviation of the Gaussian beam. The beam power flux $P$ at the point $\textbf{x}$ is given as
	\begin{equation}
		P(\textbf{x},\Theta_2) = \frac{P_0}{2 \pi \sigma^2} e^{-\frac{ (x^2+y^2)}{ 2 \sigma^2}},
	\end{equation} 
	where $P_0$ (in GW) is the total beam power. 
	In Eq.\ \eqref{eq:LIB}, the matrix $L_B^I$ relates the time derivative of the Euler angles to the body frame components of the angular velocity and derived as \cite{shirin2021modeling}. In Eq.\ \eqref{eq:force}, $R_B^I$ is the rotation matrix associated with the considered rotation convention  \cite{shirin2021modeling}.

\begin{figure}[h]
		\centering
		\includegraphics[scale = 0.35]{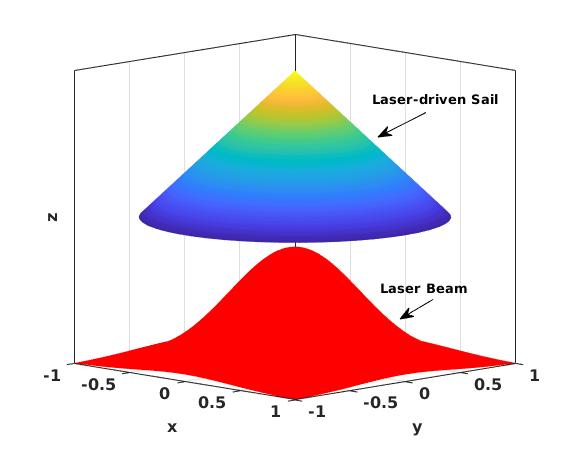}
		 	\vspace{-0.5 cm}
		\caption{Schematic illustration of a sail riding on a Gaussian laser beam.}
		\label{fig:beam}
	\end{figure}

	We now assume the material of the sail has a damping effect. Common qualified dampers may include viscous and
	magnetic induction devices and physical implementation can be obtained \cite{rafat2022self}. These effects will improve the stability performance of the beam-driven sail \cite{srivastava2019stable}. We model the damping effect in the equations of motion, and the dynamics in \eqref{eq:dynamics_c} and \eqref{eq:dynamics_d} can be modified as
	\begin{subequations}
		\begin{align}
		\dot{\textbf{v}} & = \frac{\textbf{F}}{M} + \textbf{g} - \frac{D}{M}\textbf{v}. \nonumber
%		 \\ %\tag{\ref{eq:dynamics_c}} \\
%		\dot{\boldsymbol{\omega}}_b & = J^{-1} [- \boldsymbol{\omega}^X_b \times J\boldsymbol{\omega}_b + \boldsymbol{\tau}_b]- \Gamma\boldsymbol{\omega}_b. \nonumber %\tag{\ref{eq:dynamics_d}}
		\end{align}
	\end{subequations}
	The matrix $D$ captures the damping effect and satisfies the condition: $Dij = -Dji$.
	In Fig.\ \ref{fig:beam}, a schematic illustration of a sail riding on a Gaussian beam is presented.
%
%	\begin{figure}[h]
%		\centering
%		\includegraphics[scale = 0.3]{./fig/bb.pdf}
%		\caption{A schematic representation of the sail's dynamics.}
%		\label{fig:sail-axis}
%	\end{figure}
%

\section{Levitation of the Laser Beam-driven Sail}
 	\subsection{Hypothetical Setup of the Experiment}
\begin{figure}[h]
		\begin{center}
			\includegraphics[scale = 0.65]{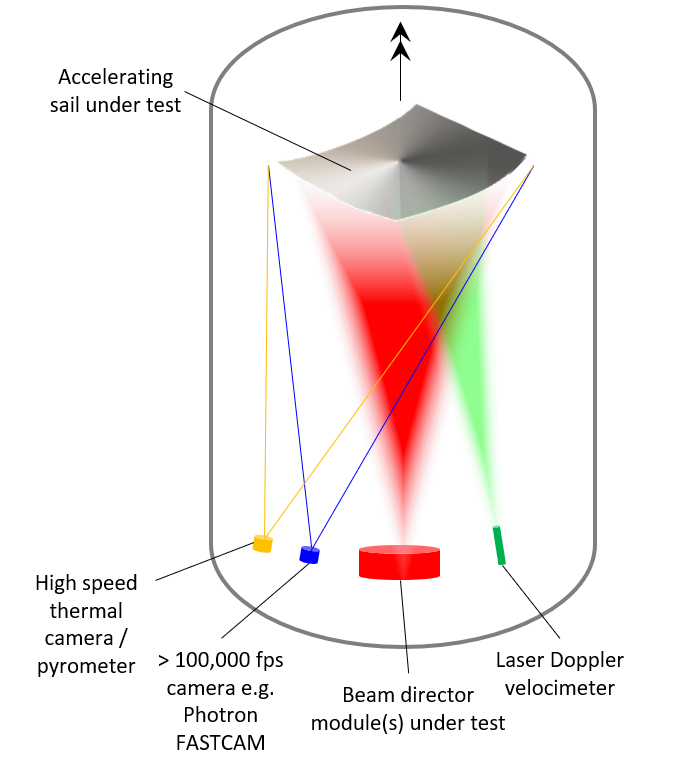}
			     	\vspace{-0.5 cm}
			\caption{Schematic of a sub-scale experimental laser-driven sail. Credit: Breakthrough Foundation.}
			\label{fig:experiment}
		\end{center}
	\end{figure}

	For an experimental setup, we assume that a sail is placed in a vacuum chamber and is driven by a Gaussian laser beam as discussed previously. We also assume that the velocometer acts like a sensor and can send the feedback signal to the actuator (laser)  to produce the force from the beam to levitate the sail. A hypothetical representation of the experiment is shown in Fig.\ \ref{fig:experiment}. 
	\subsection{Lyapunov Based Design}
	It turns out that we need to design a nonlinear closed-loop control law that not only can levitate the sail at an assigned	height but also guarantees the closed-loop stability of the sail. In this case, the force directly drives the states $z$ and $v_z$ and the other states indirectly. Thus, the states $z$ and $v_z$ are considered as actuated states. The dynamics of the actuated system can be written as
	\begin{subequations}\label{eq:1D}
		\begin{align}
			\dot{z} & = v_z,\\
			\dot{v}_z & = g + \frac{F_z}{M}, 
		\end{align}
	\end{subequations}
	where $F_z$ is the $z$ component of the force $\textbf{F}$ given in Eq.\ \eqref{eq:force}. 
	We can rewrite Eq.\ (\ref{eq:1D}) as,
	\begin{equation} 
		\begin{bmatrix}  \dot{z} \\ \dot{v}_z 
		\end{bmatrix}= 
		\left[\begin{array}{c}
			v_z \\ g 
		\end{array}\right] + \begin{bmatrix}  0 \\ 	\frac{G_z}{M} 
		\end{bmatrix}u,
	\end{equation}
	where the vector quantity,
	\begin{subequations}
		\begin{align*}
			\textbf{G} & =   \frac{1}{ c \pi \sigma^2} (R_I^B)^{-1} \iint_{S(\Theta_1)}    e^{-\frac{ (x^2+y^2)}{ 2 \sigma^2}} \hat{\textbf{b}}. \hat{\textbf{n}}(\textbf{x})  \textbf{dS}, \\
			u & = P_0,
		\end{align*}
	\end{subequations}
	and $G_z$ is the $z$ component of $\textbf{G}$. Note $F_x = G_x$, $F_y = G_y$, and $F_z = G_zu$. 
	The actuator only can produce force along the $z$-direction but not any applied  torque.  But the torque due to the force $\textbf{F}$ is still considered. If there is any angular perturbation, \textit{i.e.}, $(R_I^B)^{-1} \ne I_3$, non-zero $x$ and $y$ components of $\textbf{F}$ will be produced.
	Moreover, we assume that spinning of the sail around the $z$-axis is allowed and will not affect the dynamic stability, but the spinning around $x$ and $y$ will as these spinnings will tumble the sail. We separate the actuated and underactuated states as follows
	\begin{align*}
		\textbf{x}_a &= 	\begin{bmatrix} z & v_z	\end{bmatrix}^T,\\
		\textbf{x}_u &= \begin{bmatrix} x & y & \psi & \theta & \phi &
			v_{x} & v_{y} & \omega_x & \omega_y & \omega_z
		\end{bmatrix}^T.
	\end{align*}
	Consider the Lyapunov function for the actuated states to design a closed-loop controller
	\begin{equation} 
		V_a (\textbf{x}_a) = \frac{1}{2} z^2 + \frac{1}{2} v_z^2 
	\end{equation}
	of which the time derivative is
	\begin{align}
		\dot{V}_a  & = z v_z +v_z \left(g+\frac{G}{M} u\right).
	\end{align}
	The control input 
	\begin{equation} \label{eq:input}
		u = P_0 =  \frac{-M}{G} \left[ g +  z + v_z\right]
	\end{equation}
	leads to $\dot{V}_a = - v_z^2 <0$. If we want to levitate the sail at $z=z_d$, the transformation $e = z-z_d$ accommodates the case and we obtain
	\begin{equation} 
		u = P_0 =  \frac{-M}{G} [ g+ e + v_z].
	\end{equation}
	This feedback control law $u$ levitates the sail at $z_d$. Now we need to analyze the local stability of the internal dynamics. The internal dynamics can be written as
	%\as{
	%
	\begin{equation}\label{eq:internal}
		\begin{bmatrix}  \dot{x} \\ \dot{y}\\ \dot{\boldsymbol{\alpha}}_E \\ \dot{v}_x \\ \dot{v}_y \\\dot{\boldsymbol{\omega}}_b
		\end{bmatrix} = \underbrace{
		\begin{bmatrix}  v_x \\ v_y \\  L^I_B \boldsymbol{\omega}_b \\ 0 \\ 0 \\ J^{-1} [- \boldsymbol{\omega}_b^{X} \times J\boldsymbol{\omega}_b] 
		\end{bmatrix}}_{\bar{\textbf{f}}} + 
		\underbrace{
		\begin{bmatrix}  0 \\ 0 \\  0 \\ \frac{F_x}{M} \\  \frac{F_y}{M} \\ J^{-1} \boldsymbol{\tau}_b
		\end{bmatrix}
		}_{\bar{\textbf{g}}}.
	\end{equation}
	The states' associated with the internal dynamics can be written as,
	\begin{equation*} 
		\textbf{x}_u = \begin{bmatrix} x & y & \psi & \theta & \phi &
			v_{x} & v_{y} & \omega_x & \omega_y & \omega_z	\end{bmatrix}^T. 
	\end{equation*}

	The internal dynamics can be considered as a perturbation of a nominal system $\dot{\textbf{x}}_u(t) = \bar{\textbf{f}}(\textbf{x}_u(t))$ and the perturbed system is given in the form,
	\begin{equation}
		\dot{\textbf{x}}_u(t) = \bar{\textbf{f}}(\textbf{x}_u(t)) + \bar{\textbf{g}}( \textbf{x}_u) .
	\end{equation} 
	With the control input $u$ in Eq. \eqref{eq:input}, the full dynamics is (locally) exponentially stable. We assume that  $\bar{\textbf{g}}( \textbf{x}_u)$ is bounded on some set $\Gamma$. By considering the stability of  perturbed systems, more specifically {\it vanishing perturbation},  \cite[Section~9.1]{khalil2002nonlinear}, the internal dynamics are stable and  there exists a Lyapunov function $V (\textbf{x}_u(t))$ in $[0,\inf) \times  \Gamma$, where $\Gamma = \{\textbf{x}_u \in \mathbb{R}^{10} : \|  \textbf{x}_u\| < \bar{c}\}$ such that $\dot{V}(\textbf{x}_u) \le 0$. Here $\bar{c}$ is a positive constant. Such a set $\Gamma$ suggests that there exists an invariant subset of $\Gamma$ which can be estimated as a ROA described in the following subsection. Let 
	\begin{equation}\label{eq:lyap}
		V(\textbf{x}_u) = \textbf{x}_u^T P \textbf{x}_u(t)
	\end{equation} 
	be a Local Lyapunov Function (LLF) for the system in Eq. \eqref{eq:internal}, where $P$ is a $10\times10$ positive definite matrix. This LLF is used to compute the ROA to study the dynamic stability of the sail.
	%}

%
\section{ROA Estimation}
	To address the dynamic stability of the sail, it is sufficient to analyze the dynamic stability of the internal dynamics. Specifically, we  estimate the ROA of the internal dynamics by using Lyapunov theory and SOS programming. 	
	In SOS programming, the SOS decomposition technique is used to convert the cost and constraints such that the ultimate optimization problem becomes a Semidefinite Programming Problem (SDP) \cite{parrilo2000structured}. 
	In \cite{peet2009exponentially}, it has been proved that locally stable polynomial systems admit polynomial LLFs on compact sets. For polynomial systems and polynomial  LLFs, one can maximize the size of the ROA using SOS programming (\textit{e.g.}, \cite{prajna2005sostools,majumdar2014control}). Here the procedure to estimate the ROA is presented for a general nonlinear system of the form $\dot{\textbf{x}}(t) = \textbf{f}(\textbf{x})$.
	
	\subsection{Procedure to Estimate the ROA}
	Consider that $V(\textbf{x})$ is an LLF for the equilibrium $\textbf{x}=\textbf{0}$ of a nonlinear system $\dot{\textbf{x}}(t) = \textbf{f}(\textbf{x})$. If $\textbf{x}=\textbf{0}$ is asymptotically stable such that the following condition holds  
	\begin{equation}\label{eq:set}
		V(\textbf{x}) \le \rho \implies \dot{V}(\textbf{x}) < 0,
	\end{equation}
	then the $\rho$ sublevel set of $V(\textbf{x})$ is an inner subset of the ROA. 
	%In terms of set containment the condition in Eq.\ \eqref{eq:set} is equivalent to,
	%
%	\begin{equation}
%		\{\textbf{x} \in \mathbb{R}^n : V(\textbf{x}) \le \rho\}\backslash\{0\}\} \subseteq \{\textbf{x} \in \mathbb{R}^n : \dot{V}(\textbf{x}) < 0 \}.
%	\end{equation}
	%
	This condition can be solved by using the S-procedure \cite[Chapter~2.6.3]{boyd1994linear}. For a general polynomial, S-procedure is stated by the following lemma.
	\begin{lemma}
		Let $q_0, q_1,\cdots, q_m$ be quadratic functions on $\textbf{x} \in \mathbb{R}^n$. If it is required that   
		\begin{equation}
			q_0(\textbf{x})\ge0 \text{ such that } q_i (\textbf{x}) \ge 0, i = 1,2,\cdots,m
		\end{equation}
		then there exist polynomials $s_i(\textbf{x}) \ge 0,  i = 1,\cdots,m$ such that
		\begin{equation}
			q_0(\textbf{x}) - \sum_{i=1}^{i=m} s_i(\textbf{x})q_i(\textbf{x}) \ge 0.
		\end{equation}
		Here $s(\textbf{x})$ is a  multiplier polynomial.
	\end{lemma}	
	By using the above lemma, a sufficient condition is,
	\begin{equation}\label{eq:Non}
		p(\textbf{x})(V(\textbf{x})-\rho) + s(\textbf{x})\dot{V}(\textbf{x}) \ge 0
	\end{equation}
	to hold the condition in Eq. \eqref{eq:set} to be true. By replacing the non-negativity condition in Eq. \eqref{eq:Non} with an SOS constraint, the following SOS optimization problem can be formulated to estimate the ROA
	\begin{subequations}\label{eq:opt}
		\begin{align}
		\hspace{-2.2 in} \boldsymbol{\Max{\rho \in \mathbb{R}, s \in \mathcal{S} }} \quad & \rho \label{eq:obj} \\
		\textbf{ subject to } 	 \quad 	
			&  p(\textbf{x})(V(\textbf{x})-\rho) + s(\textbf{x})\dot{V}(\textbf{x}) \in SOS. \label{eq:const_b}
		\end{align}
	\end{subequations}
	\noindent Here $\mathcal{S}$ is a given subspace of polynomials in Eq. \eqref{eq:obj}, for example, one consisting of all quadratic polynomials or all quartic polynomials, $p(\textbf{x})$ is a deterministic positive definite polynomial, for example, $p(\textbf{x})= \textbf{x}^T\textbf{x}$. The optimization problem in Eq.\ \eqref{eq:opt} is an SDP and can be solved numerically. 	For polynomial systems and polynomial LLF one can maximize the invariant subset of the ROA by using SOS programming. Therefore, the step-by-step procedure is as follows,
	\begin{itemize}
		\item Express the nonlinear system $\dot{\textbf{x}}(t) = \textbf{f}(\textbf{x})$ in the form of polynomials  by series expansion. Although the dynamical system might not be directly	expressed in the form of polynomials, it can be approximated by series expansion  \cite{majumdar2014control}.
		\item Find an LLF $V(x)$ for the equilibrium point $\textbf{x}=\textbf{0}$ of the nonlinear system. An LLF can be found by using Eq.\ \eqref{eq:lyap} and then compute $\dot{V} = \frac{\partial V}{\partial \textbf{x}}\dot{\textbf{x}}$. 
		\item Solve the SOS optimization problem in Eq.\ \eqref{eq:opt}.
	\end{itemize}
	\subsection{SOS Programming Solver} \label{sec:solver}
	A software	package using the Systems Polynomial Optimization Toolbox (SPOT) \cite{megretski2010systems} features is used to set up the polynomial algebra and  the large-scale SOS programming arising from our problem. We use MOSEK as the optimization solver. A complete implementation is available online \footnote[1]{\url{https://github.com/anirudhamajumdar/spotless/tree/spotless_isos}}.

	\section{Numerical Results}
	\subsection{Parameter Setting for the Dynamics}
	For simulations, we consider the proposed Starshot design parameters: the beam power $P_0$ is measured in GW, the FWHM of the beam $w$ is set to the base radius of the sail, a sail with mass $m_s = 10$ g, the mass of the payload $m_p = 10$ g (if any, otherwise, $m_p = 0$). 
	The integrals in Eqs. \eqref{eq:force} and \eqref{eq:torque} were approximated by discretizing the beam into a grid of 100 × 100 rays. 
	The path of each ray is then traced as it intersects the sail and is reflected off of its surface. The net change in momentum of each ray is calculated, and the resulting forces and torques are applied to the sail. 
	For matrix $D$ we set $D_{11} = D_{12} = D_{22} = 0.01$, $D_{13} = D_{23} = D_{33} = 0.0$. 
	%
	%To compute the state trajectories, the differential equations of motion described in Section III	are integrated using the standard fourth-order Runge–Kutta method.
	%
	\subsection{Numerical Computation of the ROA}
	In this section, we numerically compute the ROA of the sail and conduct its dynamic stability analysis. For the ROA computation, first, express the internal dynamics in Eq. \eqref{eq:internal} in the form of polynomials by Taylor expansion of degree 3. 
	We find a LLF $V(\textbf{x}_u)$ as we discussed previously. Then, compute $\dot{V} = \frac{\partial V}{\partial \textbf{x}} \dot{\textbf{x}}$. The ROA can be approximated by solving the optimization problem in Eq.\ \eqref{eq:opt}. For simulations, we choose $\mathcal{S}$ as a set of quadratic polynomials. 

	\subsubsection{The ROA of a Conical Sail}
\begin{figure*}[t]
		\centering
		\includegraphics[scale = 0.68]{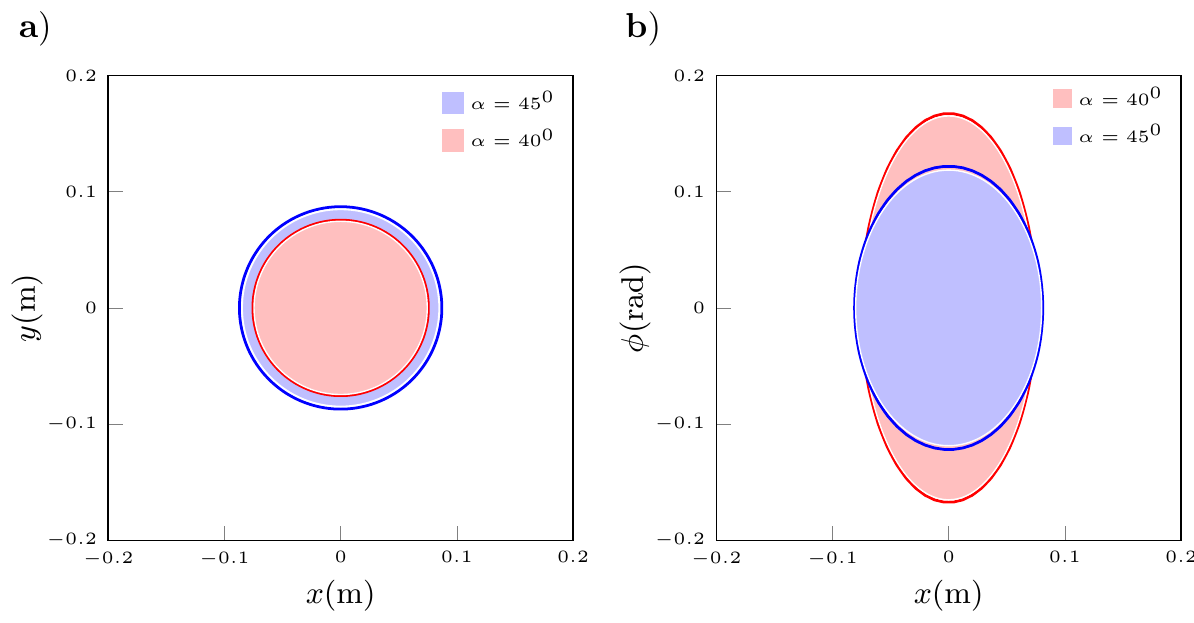}
		\includegraphics[scale = 0.68]{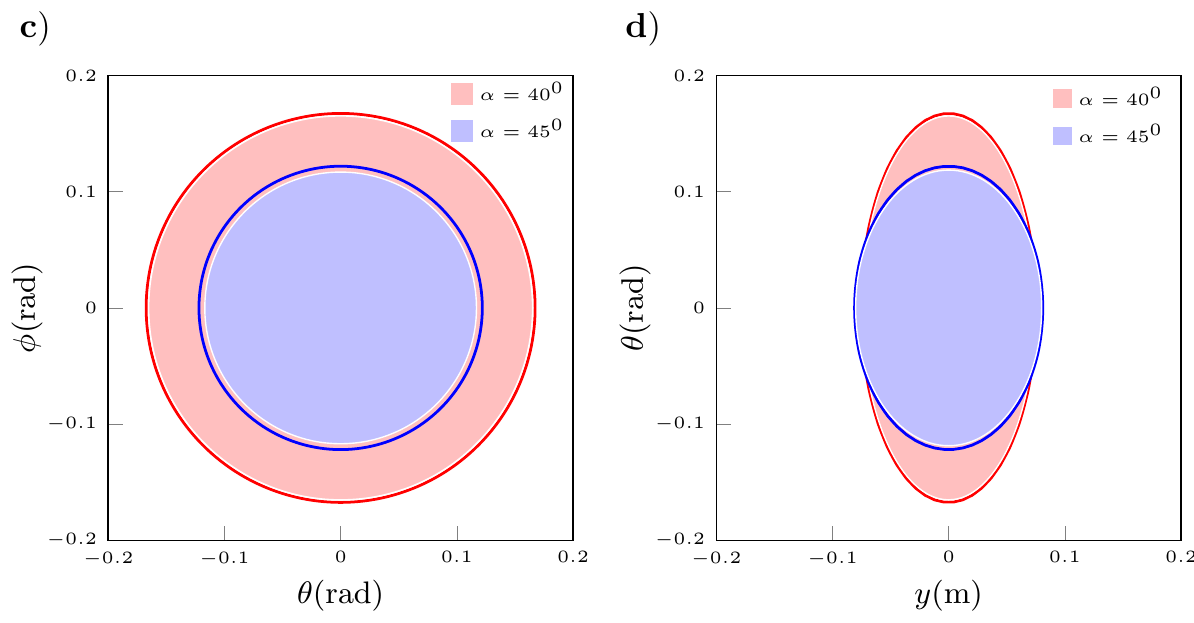}
		     	\vspace{-0.3 cm}
		\caption{The ROA of the conical sail projected in different planes (a)-(d). The pink region is associated with the cone angle $\alpha = 40^{\circ}$ and the light blue region is associated with $\alpha = 45^{\circ}$. The projected region  in the (a) $x-y$ plane, (b) $x-\phi$ plane, (c) $\theta-\phi$ plane, and (d) $y-\theta$ plane.}
		\label{fig:xy}
	\end{figure*}

	We consider a conical sail with $c_0 = R \tan(\alpha)$, $c_1 = - R \tan(\alpha)$, $c_2 = c_3 = c_4 = 0$ in Eq.\ \eqref{eq:polynomial_sail}. The base radius is set to $R = 1$, the mast length $l = 2$ m, and the levitation height $z_d=10$ m. 
	We conduct the stability analysis with respect to the cone angle $\alpha$. It is found that the $\alpha$ is one of the crucial parameters for the conical sail \cite{chahine2003dynamics}. We compute the ROA for $\alpha = 40^{\circ}$ and $\alpha = 45^{\circ}$ and plot the projected regions in the $x-y$, $x-\phi$, $y-\theta$, and $\theta-\phi$ planes. The other projections of the ROA are omitted due to symmetry or redundancy.
	In Fig.\ \ref{fig:xy}(a)-(d), we plot the projected region of the ROA on the $x-y$, $x-\phi$, $\theta-\phi$, and $y-\theta$ planes, respectively. The projected regions are shaded with colors.
	\begin{figure*}[t]
		\begin{center}
			\includegraphics[width=0.82\textwidth]{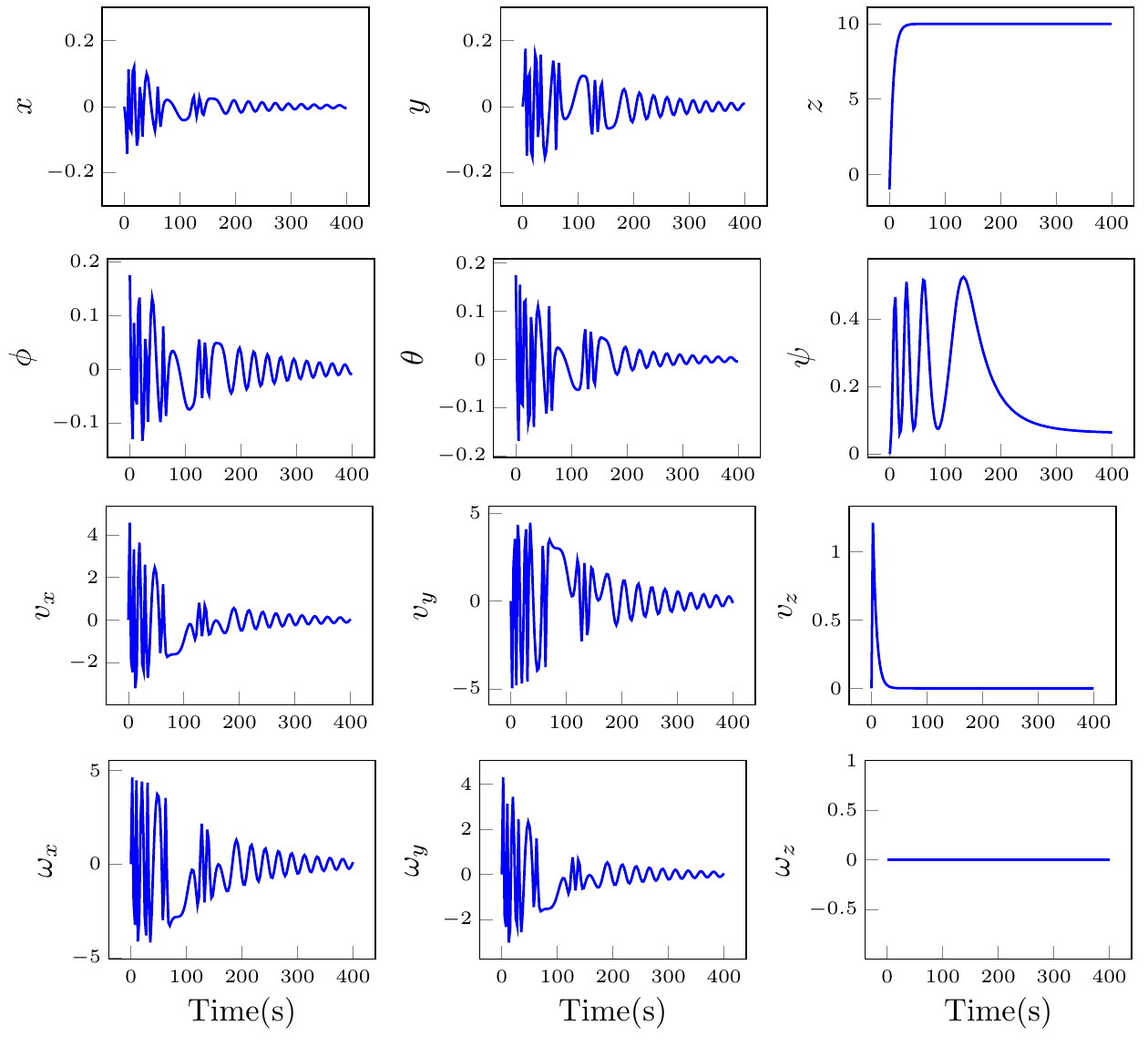}
			 	\vspace{-0.5 cm}
			\caption{State histories for the conical sail under actuation.  The cone angle of the sail is $\alpha = 40^{\circ}$ and the base radius $R = 1$ m.  The initial condition is chosen such that the roll and pitch angle perturbations are about $10^{\circ}$ (0.1745 rad).}
			\label{fig:closedcontrol_F}
		\end{center}
	\end{figure*}
	The sail with $\alpha = 40^{\circ}$ can tolerate transverse perturbations of at least 8\% of the base radius $R$ (Fig.\ \ref{fig:xy}(a)), and roll and pitch perturbations of about $10^\circ$ (0.17 rad) (Fig.\ \ref{fig:xy}(c)). The same conical sail with $\alpha = 45^{\circ}$ can tolerate transverse perturbations of at least 9\% of the base radius, and roll and pitch perturbations of about $8^\circ$ (0.13 rad).
	Upon comparison with the regions, we observe that the ROA of the conical sail with $\alpha = 40^{\circ}$ can tolerate more pitch and roll perturbations than the other one. %It is important to mention that the dynamical stability analysis suggests that the conical sail with $\alpha = 40^{\circ}$  is more acceptable for Starshot project than the other one.  
	To check the dynamic stability of the conical sail, the state trajectories of the full system in Eq. \eqref{eq:dynamics} are computed with the initial roll and pitch angles perturbed about $10^{\circ}$ (0.17 rad).
	In Fig.\ \ref{fig:closedcontrol_F}, the state time histories of the dynamics with closed-loop control $u$ are plotted for the conical sail with $\alpha = 40^{\circ}$.  The sail is levitated at 10 m. Also, in each case, the states can tolerate the initial perturbations and becomes steady after a certain amount of time. 
	\begin{figure*}[t]
		\centering
		\includegraphics[scale = 0.68]{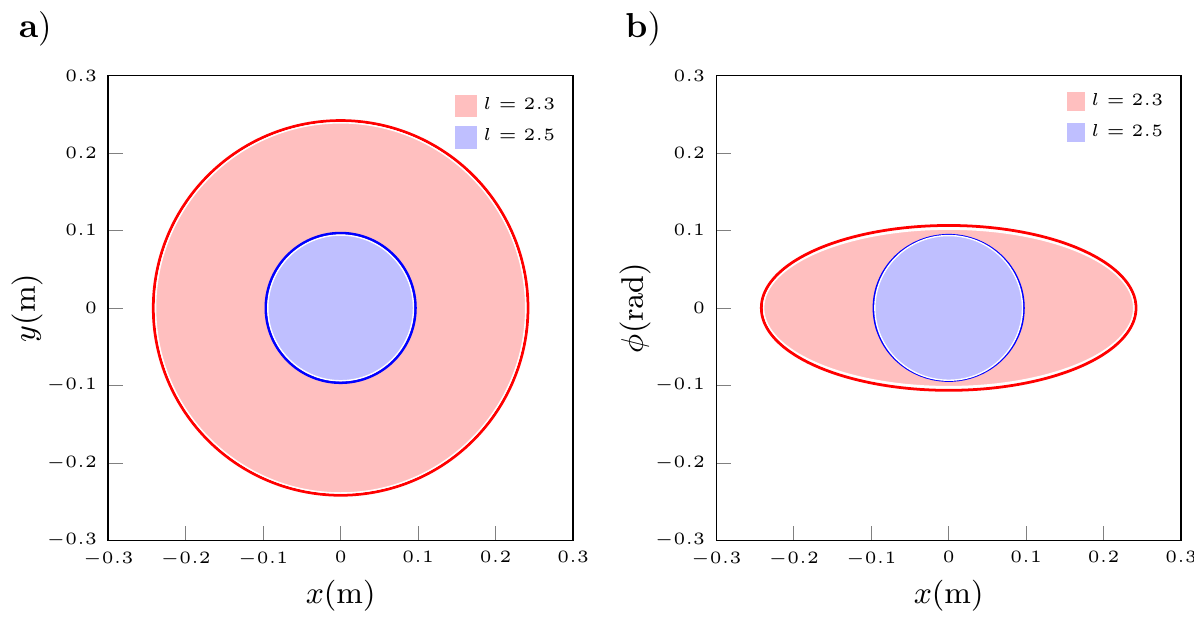}
		\includegraphics[scale = 0.68]{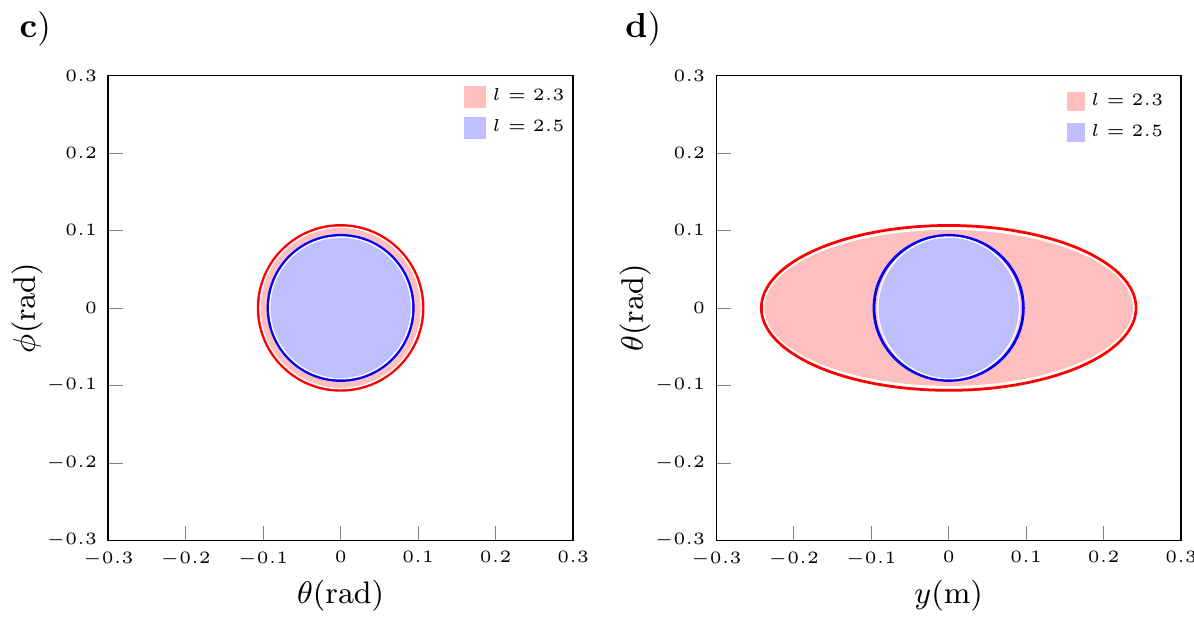}
		     	\vspace{-0.3 cm}
		\caption{The ROA of the spherical-cap sail projected in different planes (a)-(d). The pink region is associated with mast length $l = 2.3$ m and the light blue region is associated with  $l = 2.5$. The projected region  in the (a) $x-y$ plane, (b) $x-\phi$ plane, (c) $\theta-\phi$ plane, and (d) $y-\theta$ plane.}
		\label{fig:xy_sphere}
	\end{figure*}
	\begin{figure*}[t]
		\begin{center}
			\includegraphics[width=0.82\textwidth]{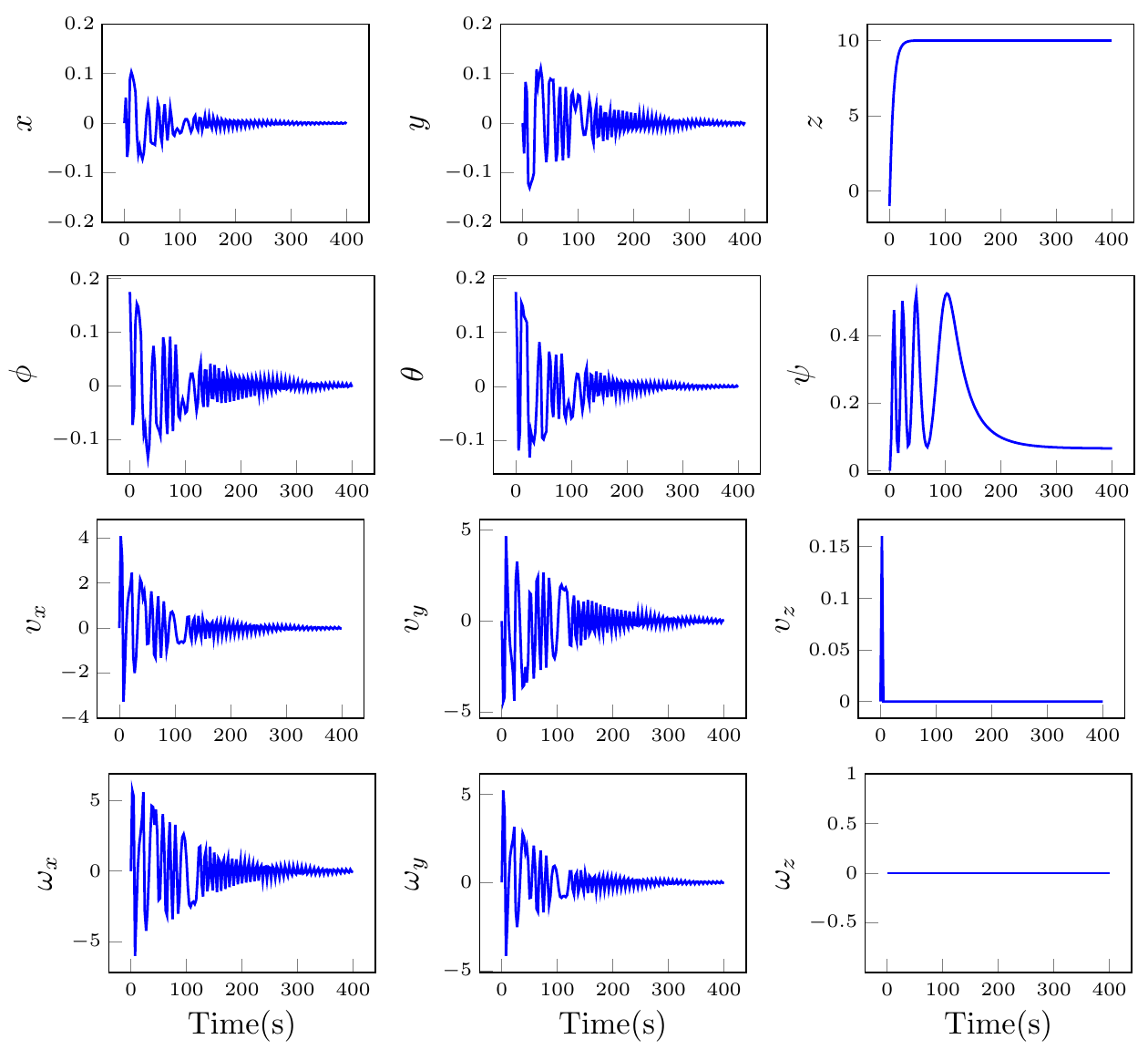}
			     	\vspace{-0.5 cm}
			\caption{State histories for the spherical-cap sail under actuation.  The length of the payload is $l = 2.3$ m and the base radius $R = 1$ m.  The initial condition is chosen such that the roll and pitch angle perturbations are about $10^{\circ}$ (0.1745 rad).}
			\label{fig:closedcontrol_F_sphere}
		\end{center}
	\end{figure*}
	
	\subsubsection{The ROA of a Spherical-cap Sail}
	We next consider a part of a spherical sail with base radius $a = 0.5$ m and with a radius of curvature $R = 1$ m  in Eq.\ \eqref{eq:spherical_sail}. The levitated  height is set at $10$ m.
	Now we analyze the stability of the spherical-cap sail. The interesting parameter, in this case, is the mast length $l$. In \cite{genta2017preliminary}, it has been shown that for a spherical-cap sail with base radius $a$ and a radius of curvature $R$, the condition to be stable is $l>2R$. 
	Now we compute the ROA for $l = 2.3$ m and $l = 2.5$ m and plot the projected region  on different planes as we did for the conical sail. It is important to mention that the spherical-cap sail with $l=2.1$ m is hardly stable.
	In Fig.\ \ref{fig:xy_sphere}(a)-(d), we plot the projected region in the $x-y$, $x-\phi$, $\theta-\phi$, and $y-\theta$ planes, respectively. 
	The sail with $l=2.3$ m can tolerate transverse perturbations of at least $20\%$ of the sail radius and roll and pitch perturbation of at least $6^{\circ}$ (0.1 rad). The same sail with $l=2.5$ m can tolerate transverse perturbations of 10\% and roll and pitch perturbations of $5^{\circ}$.
	Upon comparison with two shaded regions, it is observed that the spherical-cap sail  with $l = 2.3$ m can tolerate more transverse and angular perturbations than  the sail with $l=2.5$ m. %
	To check the dynamic stability of the spherical-cap sail, the state trajectories of the full system in Eq. \eqref{eq:dynamics} are computed with the initial roll and pitch angles perturbed of about $10^{\circ}$ (0.17 rad).
	For this computation, the differential equations of motion described in Section III	are integrated using the standard fourth-order Runge–Kutta method.
	In Fig.\ \ref{fig:closedcontrol_F_sphere}, the state time histories of the dynamics with closed-loop control are plotted for the spherical-cap sail with $l = 2.3$ m. The sail is levitated at 10 m. Also, in each case, the states can tolerate the initial perturbations and become steady after a certain amount of time. 
	\section{Discussion}
	
	The dynamics of the beam-driven sail are modeled as a rigid body. The sail shape parameterized as a sweep function has been presented to define the sail shape. 
	A Lyapunov-based closed-loop controller which levitates a sail at an assigned height, is designed. The dynamic stability analysis of a levitated sail is conducted through the estimation of the ROA. The ROA is computed using Lyapunov theory and SOS. Moreover, the ROA indicates how much perturbation  a sail can tolerate. A full simulation of the sail system with the control law confirms these regions.
	The dynamic stability analysis conveys that the system with closed-loop control is not only levitated, but also confirms its dynamic stability.   
	It is also noticeable  that with the control law, the sail is not only levitated but also stable even for large perturbations of the states.
	Upon comparison with the ROA for the conical sail with $\alpha = 40^{\circ}$ and $45^{\circ}$, this analysis suggests that the conical sail with angle $\alpha = 40^{\circ}$ performs better than the sail with $\alpha = 45^{\circ}$.
	A similar analysis is also performed for the spherical-cap sail for which  the  parameter of interest is the mast length $l$. The ROA for the spherical-cap sail with mast lengths $l = 2.3$ m and $2.5$ m. This dynamic stability analysis indicates that the spherical-cap sail with $l = 2.3$ m can tolerate more perturbations than that of $l = 2.5$ m. 
	By comparing the ROA of a conical sail with a spherical-cap sail, we conclude that in general, a conical sail can tolerate more angular perturbations while the spherical-cap sail can tolerate more transverse perturbations.

	%% Bibliography
%
	\bibliographystyle{IEEEtran}
	\bibliography{lightsail_stability.bib}

\end{document}